# First-principles studies of the electronic and magnetic structures and bonding properties of boron subnitride $B_{13}N_2$

Samir F. Matar [1,§,*], Vladimir L. Solozhenko [2]

[1] Lebanese German University (LGU), Sahel Alma, Jounieh, Lebanon
[2] LSPM–CNRS, Université Paris Nord, 93430 Villetaneuse, France

[§] *Former CNRS fellow researcher at Université de Bordeaux, ICMCB-CNRS, France*
[*] Corresponding author email: s.matar@lgu.edu.lb and abouliess@gmail.com

*Abstract*

*Rhombohedral $B_{12}$ unit is viewed as a host matrix embedding linear tri-atomic arrangements of elements (E) resulting in a relatively large family of boron-rich compounds with $B_{12}\{E$-$E$-$E\}$ generic formulation. The present work focuses on boron subnitride, $B_{13}N_2$ that we express in present context as $B_{12}\{N$-$B$-$N\}$. Within well established quantum density functional theory (DFT) a full study of its electronic properties is provided. Also linear triatomic arrangements in view of the existence in simple compounds such as sodium azide $NaN_3$, i.e. $Na^I\{N$-$N$-$N\}$ and calcium cyanamide, $Ca^{II}\{N$-$C$-$N\}$, we devised $Sc^{III}\{N$-$B$-$N\}$ to establish comparison with $B_{12}\{N$-$B$-$N\}$. $ScBN_2$ is calculated to be cohesive and possessing N-B-N isolated from $Sc^{III}$ with $d_{B-N}$ = 1.33 Å. In $B_{12}\{N$-$B$-$N\}$ an elongated $d_{B-N}$=1.43 Å is identified due to the bonding of N with one of the two $B_{12}$ boron substructures, B1 with the formation of "3B...N-B-N...3B"-like complex accompanied by a magnetic instability. Spin polarized (SP) calculations led to the onset of magnetization on central boron with $M=1\mu_B$ in a stable half-ferromagnetic ground state observed from the electronic density of states (DOS). The results are backed with total energy and calculations in both non-spin-polarized (NSP) and spin-polarized stabilizing the latter configuration over a broad range of volumes from M(V) plots. Further illustrative results are given with the charge densities (total and magnetic) and electron localization function (ELF).*

**Introduction**

Boron is an element that is intriguingly original regarding its allotropic varieties, and it has been the subject of numerous controversies since its discovery in the 19$^{th}$ century [1-4]. Boron exhibits structural complexity, electron deficiency, unusual binding situations leading to a large variety of compounds. α-rhombohedral boron has particularly caught our attention because the crystal structure offers large interstitial space to host light elements such as boron, carbon, nitrogen, oxygen, silicon, phosphorus, sulfur, arsenic, leading to the compounds: $B_4C$ or $B_{12}\{C_3\}$ [5,6]; $B_{13}C_2$ or $B_{12}\{BC_2\}$ [7]; $B_{13}N_2$ or $B_{12}\{BN_2\}$ [8-15]; $B_{12}O_2$ [2,16,17,18]; $B_{13}P_2$ [19]; $B_{12}P_2$ [20,21]; $B_{12}S$ [16]; $B_{13}As_2$ [16,22]; $B_{12}As_2$ [23].

The α-rhombohedral boron network can host one, two or three aligned interstitial atoms depicted respectively as {E1}, {E2-E2} and {E2-E1-E2}, within the empty central space of the rhombohedral unit. The interstitial atoms are located along the body diagonal of the cell as shown in Fig. 1 in the Wyckoff positions *1b* and *2c* (Table 2a). In these three atoms links {Y-X-Y} each end atom Y is bonded to three different boron icosahedra replacing two "three center" or Δ bonds. Fig. 1 shows the two kinds of boron substructures ({B1} and {B2}) and the interstitial space showing generic atoms {E2-E1-E2}. Also thin lines schematize with E2 (N in present study) the interaction with B1 substructure.

It is worthwhile noting that this type of three linear arrangement of atoms is encountered in compounds such as N-N-N in sodium azide, $NaN_3$, with short d(N-N)=1.21 Å [24] and N-C-N in calcium cyanamide, $CaCN_2$, with d(C-N)= 1.25 Å [25]. Alike α-$B_{12}$, both $NaN_3$ and $CaCN_2$ are rhombohedral with the *R-3m* space group.

In so far that experimental identification and crystal structure characterizations exist for $B_{12}\{BN_2\}$ [8,9], the present work focuses on its electronic and potential magnetic properties. The discussion will be based on the results regarding the energy-dependent physical properties as the charge density (total spins and magnetic one), the electron localization, the magnetic configuration dependent energy-volume equation of state and the site, and spin projected density of states. The properties of chemical bonding for pair interactions are also discussed qualitatively as based on the overlap populations.

## 1. Computational framework

For the search of the ground structure, geometry optimizations of the atomic positions and lattice parameters were carried out within the density functional theory (DFT) [26,27] to minimize the inter-atomic forces onto the minimum energy state. For this purpose, we used the plane-wave VASP code [28,29] using the projector augmented wave (PAW) method [29,30] for the atomic potentials. To account for the effects of exchange and correlation XC within DFT, the generalized gradient approximation (GGA) [31] was used. The conjugate-gradient algorithm [32] was used in this computational scheme to relax the atoms onto the ground state. The tetrahedron method with Blöchl et al. corrections [33] as well as a Methfessel-Paxton [34] scheme was applied for both geometry relaxation and total energy calculations. Brillouin-zone (BZ) integrals were approximated using a special **k**-point sampling of Monkhorst and Pack [35]. The optimization of the structural parameters was performed until the forces on the atoms were less than 0.02 eV/Å and all stress components less than 0.003 eV/Å$^3$. The calculations were converged at an energy cut-off of 500 eV for the plane-wave basis set concerning the **k**-point integration with a starting mesh of 6×6×6 up to 12×12×12 for best convergence and relaxation to zero strains.

Properties related with electron localization were obtained from real-space analysis of electron localization function (ELF) according to Becke and Edgecomb [36] as initially devised for Hartree–Fock calculations then adapted to DFT methods as based on the kinetic energy in which the Pauli Exclusion Principle is included by Savin et al. [37] ELF = $(1+ \chi_\sigma^2)^{-1}$ with 0 ≤ ELF ≤1, i.e. ELF is a normalized function. In this expression the ratio $\chi_\sigma = D_\sigma/D_\sigma^0$, where $D_\sigma = \tau_\sigma - ¼(\nabla\rho_\sigma)^2/\rho_\sigma$ and $D_\sigma^0 = 3/5 \ (6\pi^2)^{2/3}\rho_\sigma^{5/3}$ correspond respectively to a measure of Pauli repulsion ($D_\sigma$) of the actual system and to the free electron gas repulsion ($D_\sigma^0$), and $\tau_\sigma$ is the kinetic energy density. In the post-treatment process of the ground state electronic structures, the total charge density "CHGCAR", as well as the magnetic charge density "CHGCAR_magn" are illustrated.

From the geometry of the ground state structures NSP and SP, the electronic site and spin projected density of states (PDOS) and the properties of chemical bonding based on overlap matrix ($S_{ij}$) with the COOP criterion [38] within DFT, were obtained using full potential augmented spherical wave (ASW) method [39,40] and the GGA for the XC effects [31]. In the minimal ASW basis set, the outermost shells were chosen to represent the valence states and the matrix elements. They were constructed using partial waves up to $l_{max} + 1 = 3$ for Sc and $l_{max} + 1 = 2$ for B and N. Self-consistency was achieved when charge transfers and energy changes between two successive cycles were such as: $\Delta Q < 10^{-8}$ and $\Delta E < 10^{-6}$ eV, respectively. The BZ integrations were performed using the linear tetrahedron method within the irreducible rhombohedral wedge following Blöchl et al. scheme [33].

## 2. Calculations and discussion of the results

a- Preliminary calculations on the $B_{12}${E2-E1-E2} series

For the series of complexes based on the rhombohedral $B_{12}$ unit, a preliminary step was to examine them from valence electron count VEC on one hand, and to establish a comparative overview of the respective cohesive energies as averaged per atom, on the other hand. Parameter-free, non constrained total energies were obtained from successive self-consistent sets of calculations at an increasing number of **k**-points; then the cohesive energies are deducted from subtracting the energies of the atomic constituents. Table 1 presents the corresponding results. $B_{12}$ host has cohesive energy of -1.15 and VEC = 36, i.e. a closed shell; it is also found as a small gap insulator. Considering a starting $N_0 = 36$, $B_{12}$ will be hosting an increasing number of electrons brought by the constituents of the so-called complex entity {E2-E1-E2}. $B_{12}C_3$ add-up 12 more electrons and VEC = 48 translate a closed-shell insulator. Both are diamagnetic, but $B_{12}C_3$ is more cohesive than $B_{12}$, due to the establishing of covalent B-C bonds (cf. generic Fig. 2). The situation changes with $B_{13}C_2$ or with VEC = $N_0+11 = 47$, i.e. one electron less than in $B_{12}C_3$. $B_{12}${$BC_2$} has a hole in the valence band VB whose top is crossed at low magnitude by the Fermi level $E_F$, thus designating a weak metal.

Experimentally identified $B_{12}${$BN_2$} [8,9] shows an odd VEC number, i.e. with one unpaired electron. Calculations show that this thermodynamically stable subnitride is the most cohesive among the considered compounds. From the calculations, it is found characterized as a half-metallic magnet with a moment of 1$\mu_B$ (Bohr magneton) and its complete study is undertaken herein. From the fact the afore-mentioned simplest compound containing linear {E2-E1-E2}, the VEC of both ionic $Na^I N_3$ and $Ca^{II} CN_2$ have VEC = 16, and can be formulated as {N-N-N}$^{-1}$, and {N-C-N}$^{-2}$. Then it was reasonable to propose $Sc^{III}${$BN_2$} with a trend to {N-B-N}$^{-3}$ knowing that scandium, in spite of being trivalent element, will not behave as ionic as an alkaline (Na) or alkaline-earth (Ca) metal that possess Pauling electronegativities of $\chi \sim 1$ while $\chi(Sc) = 1.36$. Yet $ScBN_2$ can be expressed formally as "$Sc^{3+}$,{N-B-N}$^{3-}$" thus enabling for comparisons with $B_{12}${$BN_2$} regarding the electronic structure especially for the change of behavior of {N-B-N} upon embedding within $B_{12}$. Indeed the cohesive energy calculated for the proposed scandium boropernitride is high (-3.2 eV/at) as shown in Table 1.

After the overview results, we focus on $B_{12}\{BN_2\}$ herein, detailing and discussing the electronic and magnetic structures with illustrations of electron localizations and charge densities both total and magnetically resolved. The rhombohedral symmetry setting was used throughout.

b- Geometry optimization

The calculation results of geometry optimizations of $B_{12}\{BN_2\}$ and chemically related $B_{12}$ and $ScBN_2$ are given in Table 2. The latter was considered because of its ionic behavior and calculated based on the experimental crystal data of $CaCN_2$ [25].

For $B_{12}$, the results in Table 2a show a relatively good agreement between experimental and calculated values of the lattice constants and the atomic positions for the two boron atoms belonging to the two substructures. The agreement is also observed for $ScBN_2$ (Table 2b) with respect to $CaCN_2$ especially for the $x_N$ parameter. The volume is smaller due to the smaller atomic radius of Sc (1.62 Å) versus Ca (1.97 Å).

As announced above, $B_{12}\{BN_2\}$ showed a magnetic ground state with ΔE(SP-NSP) = -0.51 eV and an integer moment with M=1 $\mu_B$ with a half metallic behavior as further detailed hereunder. Also the calculated crystal data showed close values between the two magnetic configurations. In Table 2c only the NSP values are confronted with the experiment. In spite of some expected deviations for the lattice parameters and the internal positions parameters, the experimental structure is reasonably reproduced by the calculations. This permits exploiting the results further to describe details of electron structure.

c- Charge and magnetic densities and electron localization

i) $ScBN_2$

The illustration of the results is given by the representations in Fig. 2. In both subfigures, oppositely to Sc, electron localization is shown only at N-B-N. The ELF's show terminal N capped with non bonding electrons in Fig. 2a, in the shape of yellow volumes. Yet, electron localization is shown between B and N, signaling the B-N bonds.

Further illustration is given with the charge density at Fig. 2b with yellow envelopes characterizing the two terminal nitrogen atoms having pear-like shape flattened toward the non bonding region and sharp towards boron. This complies with the larger electronegativity of N ($\chi$~3) versus B ($\chi$~2).

ii) $B_{12}\{BN_2\}$

$B_{12}\{BN_2\}$ ELF and charge densities are presented at Fig. 3. Fig. 3a shows the ELF yellow envelopes scattered around B1 and B2 of $B_{12}$ host, thus signaling the cohesion. But also major features appear around the central N-B-N where ELF volumes are seen between N and their B1 neighbors, oppositely to the capped ELF at N in $ScBN_2$. Also the localization between B and N is shown as in $ScBN_2$, whence the bonded N-B-N. The major effect of the interaction of N with B atoms of the B1 substructure is the elongation of the B-N bond from 1.33 Å in $ScBN_2$ to 1.43 Å in $B_{13}N_2$.

Turning to the charge density projections, oppositely to $ScBN_2$, the volumes are now pointing alike in the ELF towards B1 substructures. Then embedding N-B-N into $B_{12}$ as expressed in the formula $B_{12}\{BN_2\}$ is illustrative of a bonded chemical system. This is further shown in the DOS and COOP.

In as far as a magnetic ground state was identified by the calculations with M=1 $\mu_B$ complying with VEC($B_{12}$) = 36 and the VEC($B_{13}N_2$) = 36+13 = 49, i.e. and odd number leaving on unpaired electron, and the expectation of a paramagnetic or magnetic order. Fig. 3c shows the calculated magnetic charge density projected on the atomic constituents. Clearly it is centered on central boron in N-B-N, with the remarkable 3D feature of a torus. Also small blue volumes are seen on terminal N's, pointing towards B and letting suggest some induced magnetic polarization on N.

d- Volume change of the magnetization

At this point and in so far that $B_{13}N_2$ was synthesized at high pressure [9,10], it becomes relevant to examine the volume change of the magnetization. Fig. 4 shows the calculated points (the line joining the calculated dots is only a guide for the eye). At large volumes, the magnetization saturates linearly at 1 $\mu_B$, but below ~95 Å$^3$ it drops rapidly down to zero. To some extent the sample can be used as a gauge. These observations show that the magnetic state is stable over a wide volume range. Nevertheless, it needs to be mentioned that the calculations within DFT are at zero Kelvin. Since $\Delta G = \Delta H - T\Delta S$, then $T\Delta S = 0$ and the free energy ΔG corresponds to ΔH, i.e. the enthalpy. Experimentally, the thermal effects are likely to play a significant role in the magnetic properties, i.e. a passage from ferromagnetic order to paramagnetic state. Measurements at low temperatures are likely to bring further assessment.

e- Electronic density of states DOS

Using the calculated data in Table 2, further calculations were carried out to fully describe the role of each constituent in the electronic structure. For the purpose we used the full potential ASW method [40] implementing the COOP [38]. In a nutshell the COOP are the overlap integral ($S_{ij}$) weighted DOS, they also have the unit of inverse energy 1/eV. Also along the y-axis, positive, negative and zero intensity magnitudes translate bonding, anti-bonding, and nonbonding interactions.

i) $ScBN_2$

The properties of the electronic structure and of chemical bonding inferred from the site projected density of states PDOS and the COOP for pair interactions, are shown in Figs. 5a and 5b, respectively. Along the x-axis the energy is counted as with respect to the top of the valence band VB, separated from the conduction band (CB) by an energy gap of ~2eV. The chemical system is then predicted to be insulator as proposed above from the VEC. The VB shows two distinct regions, at low energy, around -15 eV, nitrogen s states are dominant with a large peak. Nevertheless B s-PDOS as well as a small contribution from Sc-s are observed at the same energy but with much lower magnitudes, signaling quantum mixing between them, especially for B-N and less for Sc-N. Over 6 eV below $E_V$, a broad DOS block correspond to p-states arising mainly from N and less to B, a non negligible contribution from Sc p states is observed near the top of VB. The similarity of the respective three constituents PDOS's signal the quantum mixing between them and since p states are directional versus spherical s states, one expect the major bonding through the $S_{ij}$ overlap integral to occur within this p-DOS block. Lastly within the empty CB, major contribution is from Sc empty states since the electron transfer was observed from above to be towards the linear N-B-N.

Fig. 5b showing the COOP illustrates further the DOS discussion. The COOP's within the VB are all positive meaning that the chemical system is of bonding character. By showing small s-COOP contribution at low energy, the s-like bonding is not significant versus two p-bonding COOP's blocks: major N-B-N contribution and less intensive Sc-N while Sc-B interactions are negligible. These observations clearly show the N-B-N as the major stabilizing entity of $ScBN_2$.

ii) $B_{12}\{BN_2\}$

*Non spin-polarized NSP calculations.*

Considering firstly NSP configuration, the site projected DOS (PDOS) and the COOP for pair interactions, are shown in Figs. 6a and 6b, respectively. Oppositely to insulating $ScBN_2$, along the x-axis the zero energy is now at the Fermi level $E_F$. Within the valence band VB, the DOS exhibit three main energy regions: in the lowest energy part N(2s) states are found split into two narrow PDOS; the lower energy peak is due to the s-like B-N quantum mixing, while at -20 eV one finds the mixing with N and with the less electronegative B1 s- states. The energy region {-15; -1 eV} is mainly dominant with p states showing B1-B2 quantum mixing as well as B1-N-through the p states in the lower energy part. Low-intensity B-2p PDOSs are seen. However, a most interesting feature appears at the Fermi level crossed by a relatively large B PDOS; underneath much smaller intensity PDOS are also seen arising from N and B1. Such high-intensity DOS@$E_F$ signals instability of the electron system; i.e. B-2p in such total spins NSP configuration [41]. Lowering of the energy is expected upon accounting for two spin channels, i.e., in spin-polarized SP calculations which were done subsequently. Above $E_F$ the empty conduction band CB mirrors the two features of N-N and B-B quantum mixings observed below $E_F$ with the VB.

Fig. 6b shows the COOP for the different interactions. The major binding region is expectedly in the VB characterized by the p states because such states are *x,y,z* -oriented whereas *s* states are spherical as mentioned above in $ScBN_2$ discussion. The bonding follows relatively the PDOS's allure, i.e. with B-N and B1-N bonding character whereas B2-N COOP's are mainly anti-bonding due to their large separation. B1-N COOP's start to be anti-bonding in the higher energy part {-8; -1 eV} of the valence band; where B1-B2 bind to ensure for the stability of the $B_{12}$ skeleton. At $E_F$ anti-bonding (negative) COOP signals the instability of the electronic system in such spin degenerate configuration. From these observations showing bonding N-B-N and B1-N arising from the linear complex besides $B_{12}$'s B1-B2 the higher cohesion of $B_{12}\{BN_2\}$ versus $B_{12}$ can be understood.

*Spin-polarized SP calculations.*

The NSP DOS exhibiting the peculiar feature of a large B-PDOS at $E_F$, is indicative of the instability of the electronic system in such spin degenerate configuration. SP calculations considering two spin channels ↑ and ↓ resulted indeed in a more stable state with ΔE(SP-NSP) = –0.64 eV, a magnitude slightly larger than with geometry optimization procedure, but close enough to cast confidence on the results. The site and spin projected DOS in Fig. 7a selectively for the N-B-N sites, in the p-block energy range, and where B undergoes spin polarization. The DOSs are now distributed in two panels for ↑ and ↓ spin populations. The former labeled 'majority spins' (larger population) are shown with a fat↑, oppositely to 'minority spins' ↓ (smaller population). The difference majority – minority provides the magnetic moment, with 1 $\mu_B$. While there is no energy shift for N-p ↑, ↓ states and almost none for B underneath the N-PDOS, the remarkable feature is observed at $E_F$ crossed by large B ↑ PDOS while a 2 eV energy gap is observed for ↓ DOS, leaving the B↓ PDOS within the empty CB. The magnetic situation already known to occur in the rare ferromagnetic oxide at room temperature $CrO_2$ used in high density magnetic recording [42].

Fig. 7b shows for N-B-N the corresponding change of the bonding upon accounting for the two spin populations ↑ and ↓. As for the SP-DOS, we note the general energy down-shift for spin-up↑ and up-shift for spin-down↓, mostly observed at $E_F$ and less within the p-block in the {-15 – -2 eV}. Indeed the top of the valence band is dominated by anti-bonding ↑ COOP (negative) relevant to the p states of boron B carrying the magnetic moment. Their counterpart COOP (green peak) is found in the CB.

## 4- Conclusions

Based on starting experimental observations, original properties arising from rhombohedral $B_{12}$ unit with triatomic linear {E:E:E} complex-related have been shown for boron subnitride $B_{12}\{BN_2\}$. Firstly, based on DFT quantum calculations of the cohesive energies, enhanced cohesion of $B_{12}$ embedding {E:E:E} for a series of known compounds was observed. Specifically, experimentally evidenced $B_{12}\{BN_2\}$ was found characterized as cohesive with a magnetic ground state where a magnetic polarization develops on boron in an elongated linear N-B-N ($d_{B-N}$ = 1.43 Å) versus $d_{B-N}$ = 1.33 Å in the ionic $ScBN_2$ devised and calculated herein. The binding between N and one of the two boron substructures exhibiting a "3B…N-B-N…3B"-like complex, is illustrated from charge density and ELF. A magnetic ground state with a moment of 1 $\mu_B$ carried by boron and a half metallic ferromagnetic behavior characterizes $B_{13}N_2$ over a broad volume range.

Table 1. Cohesive energies of *rh*-$B_{12}$ based compounds with *R-3m* space group highlighting the embedded triatomic linear complex, with generic $B_{12}\{E_2$-$E_1$-$E_2\}$ formula. VEC = valence electron count. Energies are in units of eV.

| System | $E_{Tot.}$ | $E_{coh}$/at | VEC | Magnetic state |
|---|---|---|---|---|
| $B_{12}$ | -80.49 | -1.15 | $N_0$=36 | diamagn. / insulator |
| $B_{12}\{BC_2\}$ | -106.35 | -1.41 | $N_0$+11 | hole-paramagn. / weak metal |
| $B_{12}\{C_3\}$ | -108.80 | -1.51 | $N_0$+12 | diamagn. / insulator |
| $B_{12}\{BN_2\}$ | -106.59 | -1.61 | $N_0$+13 | paramagnet / magn. order? |
| *ScBN$_2$* | *-35.61* | *-3.51* | *13+3 = 16* | *insulator* |

**Notes:** $B\ (2s^2\ 2p^1) \rightarrow VEC=3$; $C\ (2s^2\ 2p^2) \rightarrow VEC=4$; $N\ (2s^2\ 2p^3) \rightarrow VEC=5$
*Energies of atomic constituents (eV):* $E_C$ =-6.48; $E_B$= -5.56; $E_N$=-5.11; $E_{Sc}$= -5.45

*ScBN$_2$* $E_{Tot.}$= -35.61 eV, $E_{coh.}$/at.=-3.2eV/at.

N.B. *Ca(CN$_2$)*: $E_{coh.}$/at.= -3.51 eV/at

Table 2. Experimental and calculated crystal data of rhombohedral compounds under consideration. Space group *R-3m*, N°166. Distances are expressed in units of Å, 1 Å=10$^{-10}$m.

a) $B_{12}$ [3]. $a_{rh}$= 5.057 (4.98) Å; α = 58.06° (58.47°); $V_{rh}$ = 87.378 (85.44) Å$^3$

| Atom | Wyckoff | x | y | z |
|------|---------|---|---|---|
| $B_1$ | 6h | 0.221 (0.223) | x | -0.368 (0.371) |
| $B_2$ | 6h | 0.010 (0.011) | x | -0.343 (0.347) |
| E1 | 1b | ½ | ½ | ½ |
| E2 | 2c | x | x | x |

$E_1$ and $E_2$ designate the positions of interstitial sites where additional atoms are hosted (cf. Fig. 2 and Tables 1 and 2).
d(B1-B1) = 1.65 Å; d(B2-B2) =1.77 Å.

b) $CaCN_2$ (space group *R-3m*, N°166) [25] and calculated rhombohedral $ScBN_2$.
$a_{rh}$= 5.35 (5.30) Å; α = 40.47° (36.90°); $V_{rh}$= 58.18 (48.36µ) Å$^3$

| Atom | Wyckoff | x | y | z |
|------|---------|---|---|---|
| Ca(Sc) | 1a | 0 | 0………… | 0 |
| C (B) | 1b | ½ | ½ | ½ |
| N (N) | 2c | 0.415 (0.410) | x | x |

d(Ca-N) =2.46 Å; d(C-N) =1.23 Å.
d(Sc-N) =2.24 Å; d(B-N) =1.33 Å.

c) $B_{13}N_2$ [9]. $a_{rh}$ = 5.157 (5.211) Å; α =63.73°(63.45)

| Atom | Wyckoff | x | y | z |
|------|---------|---|---|---|
| $B_1$ | 6h | 0.208 (0.197) | x | 0.692 (0.680) |
| $B_2$ | 6h | 0.003 (0.004) | x | 0.357 (0.331) |
| B | 1b | ½ | ½ | ½ |
| N | 2c | 0.383 (0.385) | x | x |

Calculated shortest distances
d(B-N)=1.43 Å
d(B1-N)=1.56 Å
d(B1-B1)=1.74 Å
d(B2-B2) =1.72 Å.
d(B1-B2) =1.80 Å.

SP(NSP) distances in Å.
d(C-N)=1. 38 (1.37); d(N-B1) =1.58 (1.58); d(B1-B1)=1.74(1.73); d(B2-B2) =1.72 (1.71); d(B1-B2) =1.79 (1.79).

*N.B. In hexagonal setup: $a_{hex.}$=5.467, $c_{hex.}$=12.3854. C (0,0, ½); N (0,0, 0.388); B1 (-0.11, -0.22, 0.112); B2 (-0.161667, -0.323333, 0.3576677)*

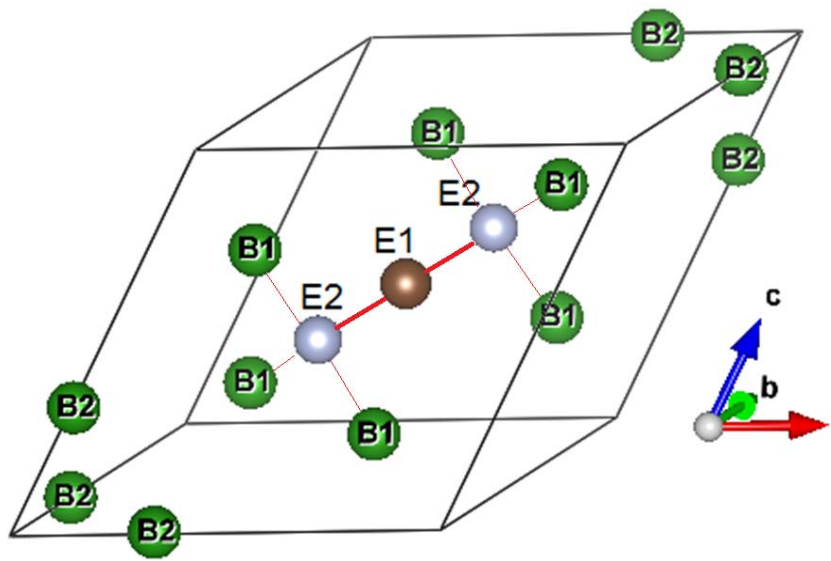

**Figure 1** Rhombohedral B$_{12}$ unit highlighting the two kinds of boron substructures and the interstitial space with the generic atoms E1 and E2 with the interaction of E2 (N in present study) with B1 substructure is schematized

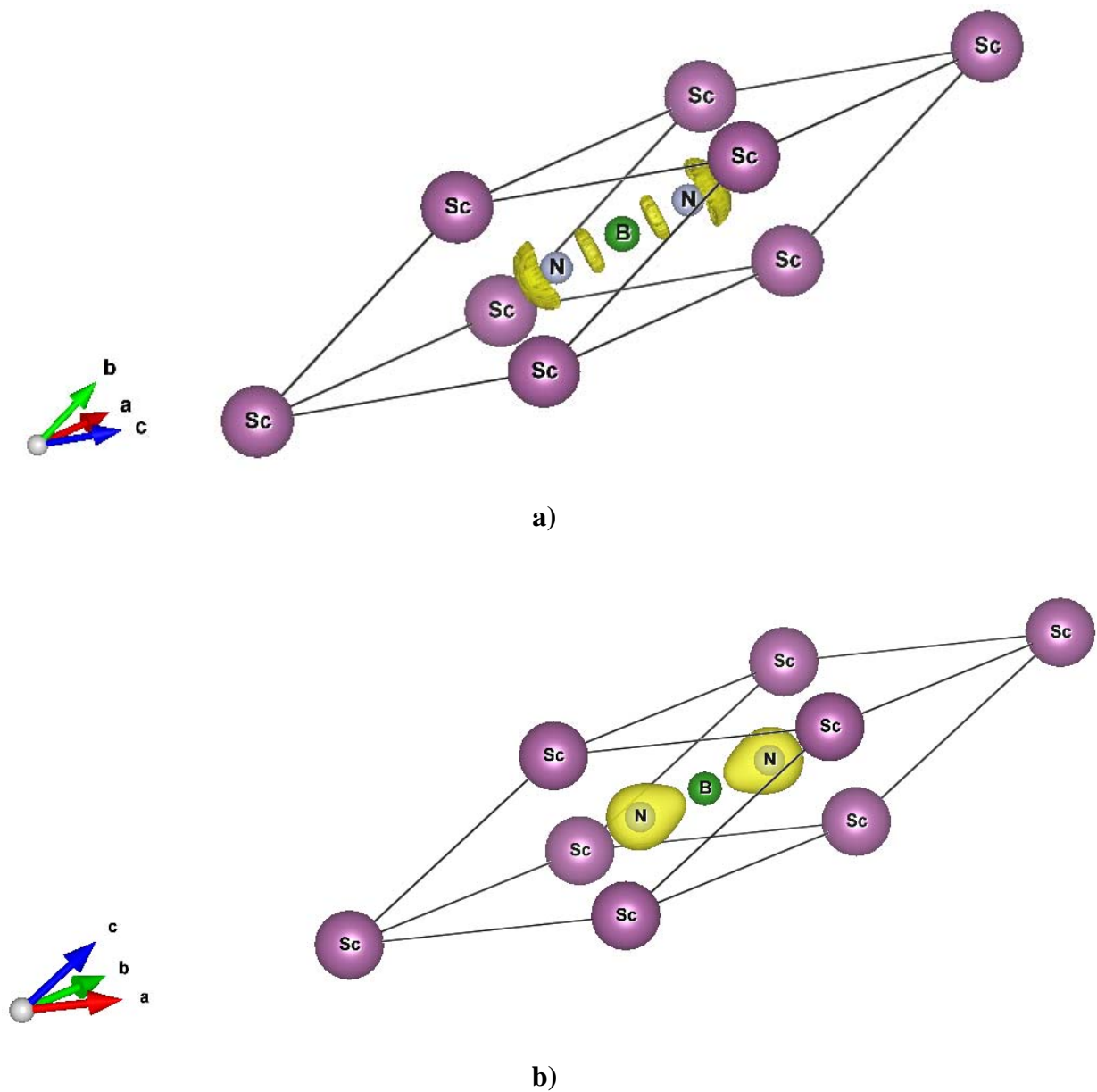

**Figure 2** Model scandium boropernitride Sc(BN$_2$) derived from calcium cyanamide Ca(CN$_2$). a) ELF 3D envelopes along the linear N-B-N; b) Charge density observed around terminal nitrogen atoms.

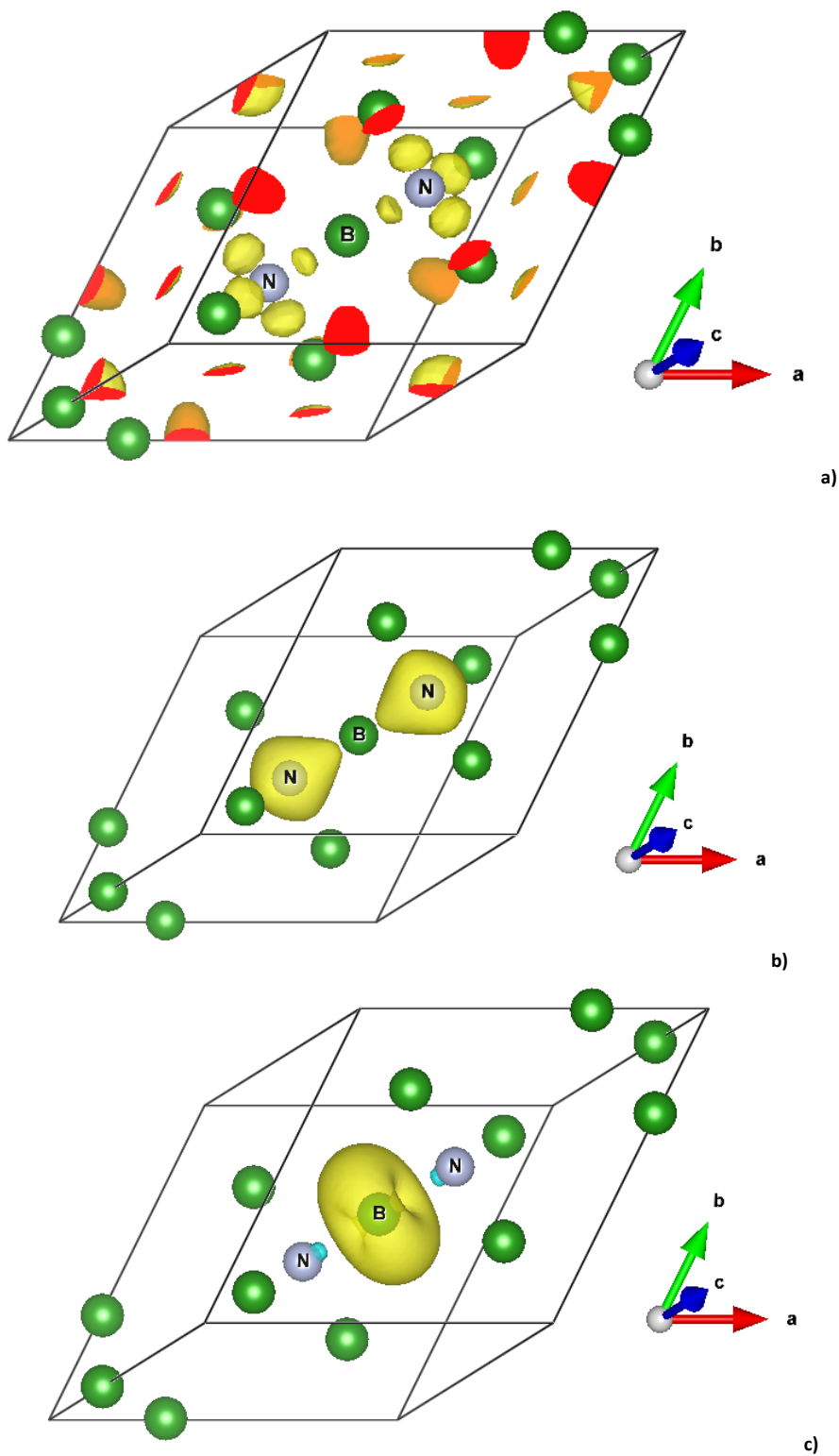

**Figure 3** $B_{13}N_2$ or $B_{12}\{BN_2\}$ stressing the central linear N-B-N. a) ELF yellow envelopes of electron localization characterizing the interactions of terminal N with 3B1 atoms. b) Total charge density shown to prevail around nitrogen and developing towards B1 atoms. c) Magnetic charge density (yellow torus) around central boron with a magnitude of 1 $\mu_B$.

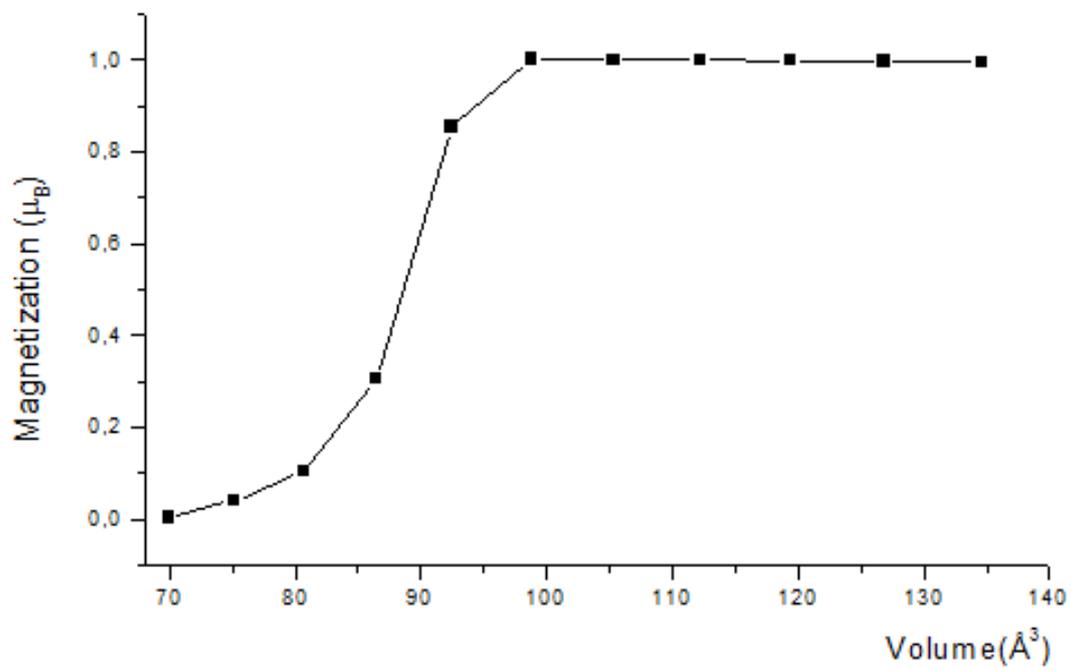

**Figure 4**  Volume change of the magnetization in $B_{13}N_2$ showing the stability of magnetic polarization over a broad volume range.

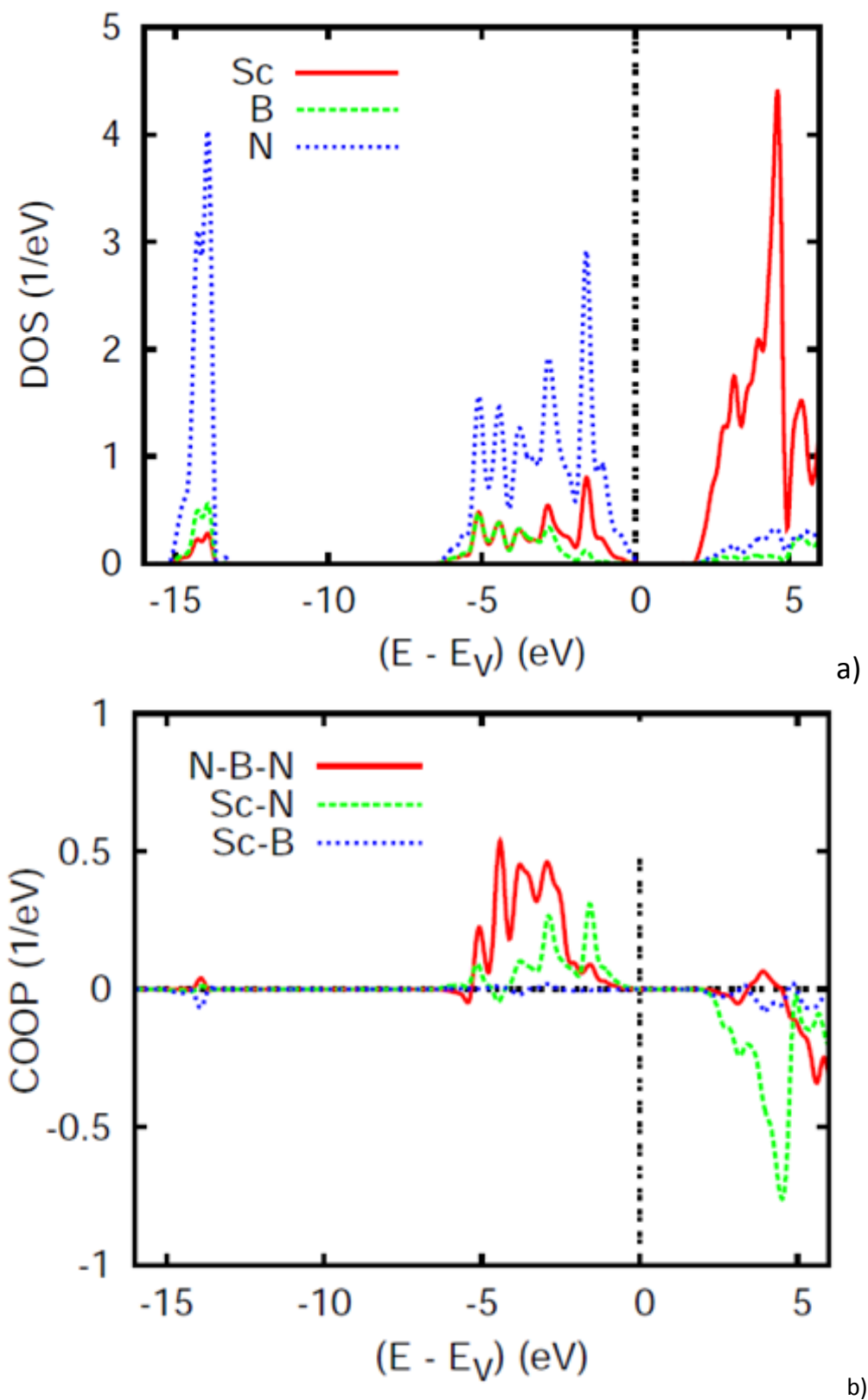

**Figure 5** ScBN$_2$. a) Site projected electronic density and b) chemical bonding.

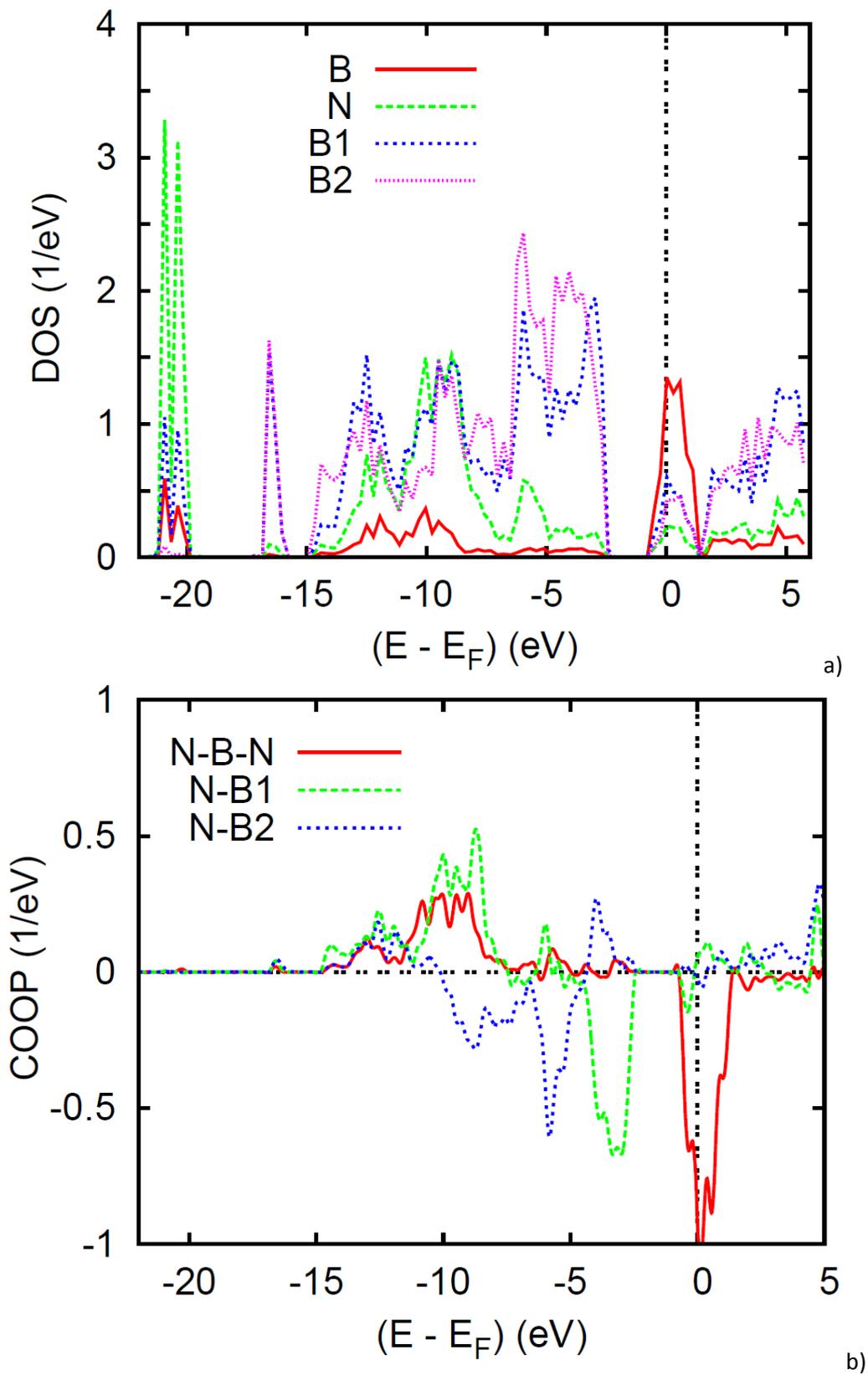

**Figure 6** $B_{13}N_2$. a) Site projected electronic density and b) chemical bonding.

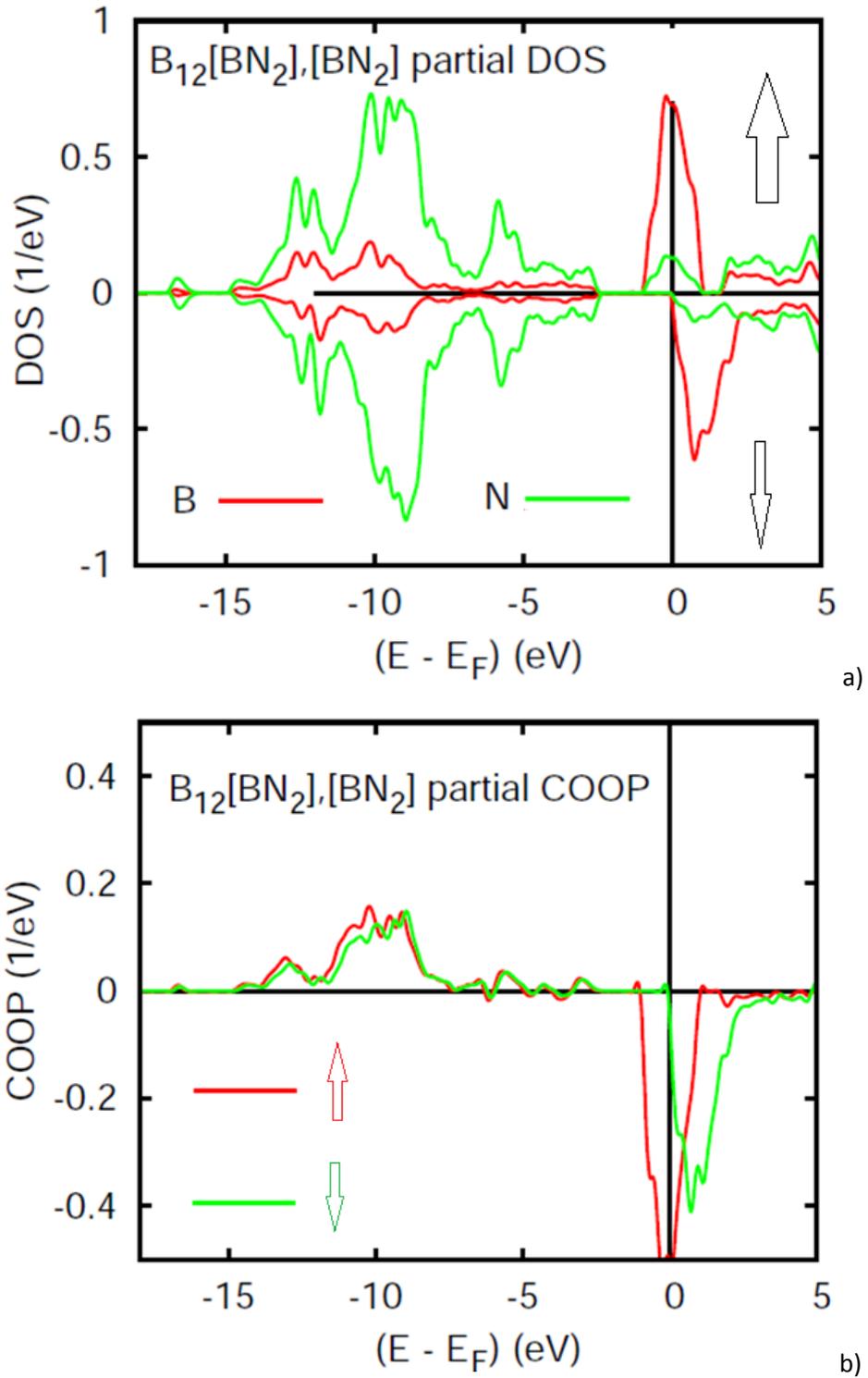

**Figure 7** $B_{13}N_2$ in the magnetic state. a) Site and spin projected density of states DOS within linear N-B-N showing for half metallic ferromagnetic behavior characterized by integer 1 $\mu_B$ moment; b) Spin projected COOP.